\documentclass[aps,prb,preprint,superscriptaddress]{revtex4}
\usepackage{graphicx}
\usepackage{amssymb}
\usepackage{amsmath}
\usepackage{color}

\begin{document}

\title{Local and collective magnetism of EuFe$_{2}$As$_{2}$}

\author{Jonathan Pelliciari}
\email[]{jonathan.pelliciari@gmail.com}
\altaffiliation[Present address: ]{Department of Physics, Massachusetts Institute of Technology, Cambridge, MA 02139, USA}
\affiliation{Research Department of Synchrotron Radiation and Nanotechnology, Paul Scherrer Institut, CH-5232 Villigen PSI, Switzerland}
\author{Kenji Ishii}
\affiliation{Synchrotron Radiation Research Center, National Institutes for Quantum and Radiological Science and Technology, Sayo, Hyogo 679-5148, Japan}   
\author{Marcus Dantz}
\affiliation{Research Department of Synchrotron Radiation and Nanotechnology, Paul Scherrer Institut, CH-5232 Villigen PSI, Switzerland}
\author{Xingye Lu}
\affiliation{Research Department of Synchrotron Radiation and Nanotechnology, Paul Scherrer Institut, CH-5232 Villigen PSI, Switzerland}
\author{Daniel McNally}
\affiliation{Research Department of Synchrotron Radiation and Nanotechnology, Paul Scherrer Institut, CH-5232 Villigen PSI, Switzerland}
\author{Vladimir N. Strocov}
\affiliation{Research Department of Synchrotron Radiation and Nanotechnology, Paul Scherrer Institut, CH-5232 Villigen PSI, Switzerland}
\author{Lingyi Xing}
\affiliation{Beijing National Lab for Condensed Matter Physics, Institute of Physics, Chinese Academy of Sciences, Beijing 100190, China}
\author{Xiancheng Wang}
\affiliation{Beijing National Lab for Condensed Matter Physics, Institute of Physics, Chinese Academy of Sciences, Beijing 100190, China}
\author{Changqing Jin}
\affiliation{Beijing National Lab for Condensed Matter Physics, Institute of Physics, Chinese Academy of Sciences, Beijing 100190, China}
\affiliation{Collaborative Innovation Center for Quantum Matters, Beijing, China}
\author{Hirale S. Jeevan}
\affiliation{Experimental Physics VI, Center for Electronic Correlations and Magnetism, University of Augsburg, 86159 Augsburg, Germany}
\author{Philipp Gegenwart}
\affiliation{Experimental Physics VI, Center for Electronic Correlations and Magnetism, University of Augsburg, 86159 Augsburg, Germany}
\author{Thorsten Schmitt}
\email[]{thorsten.schmitt@psi.ch}
\affiliation{Research Department of Synchrotron Radiation and Nanotechnology, Paul Scherrer Institut, CH-5232 Villigen PSI, Switzerland}

\date{\today}

\begin{abstract}
We present an experimental study of the local and collective magnetism of $\mathrm{EuFe_2As_2}$, that is isostructural with the high temperature superconductor parent compound $\mathrm{BaFe_2As_2}$. In contrast to $\mathrm{BaFe_2As_2}$, where only Fe spins order, $\mathrm{EuFe_2As_2}$ has an additional magnetic transition below 20~K due to the ordering of the Eu$^{2+}$ spins ($J =7/2$, with $L=0$ and $S=7/2$) in an A-type antiferromagnetic texture (ferromagnetic layers stacked antiferromagnetically). This may potentially affect the FeAs layer and its local and correlated magnetism. Fe-K$_\beta$ x-ray emission experiments on $\mathrm{EuFe_2As_2}$ single crystals reveal a local magnetic moment of 1.3$\pm0.15~\mu_B$ at 15~K that slightly increases to 1.45$\pm0.15~\mu_B$ at 300~K. Resonant inelastic x-ray scattering (RIXS) experiments performed on the same crystals show dispersive broad (in energy) magnetic excitations along $\mathrm{(0, 0)\rightarrow(1, 0)}$ and $\mathrm{(0, 0)\rightarrow(1, 1)}$ with a bandwidth on the order of 170-180 meV. These results on local and collective magnetism are in line with other parent compounds of the $\mathrm{AFe_2As_2}$ series ($A=$ Ba, Ca, and Sr), especially the well characterized $\mathrm{BaFe_2As_2}$. Thus, our experiments lead us to the conclusion that the effect of the high magnetic moment of Eu on the magnitude of both Fe local magnetic moment and spin excitations is small and confined to low energy excitations.
\end{abstract}

\maketitle
\section{Introduction} 
High temperature superconductivity (SC) was discovered in 2008 in Fe pnictides \cite{kamihara_iron-based_2008}, prompting detailed investigations of their structural, electronic, and magnetic properties \cite{stewart_superconductivity_2011,johnston_puzzle_2010,johnson_iron-based_2015,hosono_iron-based_2015}. In particular the 122 series, with the stoichiometry $\mathrm{AFe_2As_2}$ ($A=$ Ba, Ca, and Sr), have attracted a lot of attention because of the availability of relatively large single crystals ($>1$ mm$^2$), and the possibility to achieve SC by means of electron, hole, and isovalent doping \cite{stewart_superconductivity_2011, johnston_puzzle_2010,johnson_iron-based_2015, hosono_iron-based_2015}. The canonical parent compound $\mathrm{BaFe_2As_2}$ has a $\mathrm{ThCr_2Si_2}$-type structure and undergoes a tetragonal-to-orthorombic phase transition followed by collinear antiferromagnetic (AF) ordering below 140~K \cite{stewart_superconductivity_2011,johnson_iron-based_2015,hosono_iron-based_2015,johnston_puzzle_2010}. The presence of an AF phase close to SC is similar to other unconventional superconductors such as the cuprates and heavy fermions \cite{scalapino_common_2012,stewart_superconductivity_2011,johnson_iron-based_2015,hosono_iron-based_2015,fujita_progress_2011,inosov_spin_????,tranquada_superconductivity_2014, chubukov_iron-based_2015, dai_magnetism_2012,dai_antiferromagnetic_2015,lumsden_magnetism_2010,norman_challenge_2011,norman_unconventional_2013}, and has allowed to extend theoretical magnetic pairing scenarios, previously proposed, to Fe pnictides \cite{scalapino_common_2012,chubukov_iron-based_2015,das_intermediate_2014}. Therefore, the intimate interplay between antiferromagnetism and SC highlights the importance of a complete characterization of both static and dynamic magnetism in the Fe pnictides. 
The ordered magnetic moment detected in $\mathrm{BaFe_2As_2}$ is $\approx$1 $\mu_\text{B}$ \cite{dai_antiferromagnetic_2015,lumsden_magnetism_2010,stewart_superconductivity_2011,johnson_iron-based_2015,johnston_puzzle_2010}. However quantum fluctuations due to the metallicity of $\mathrm{BaFe_2As_2}$, and Fe pnictides in general, hinder the determination of the value of the fluctuating magnetic moment in slow probes \cite{mannella_magnetic_2014,vilmercati_itinerant_2012} (such as neutron scattering, NMR, and muon relaxation spectroscopy \cite{mannella_magnetic_2014, su_antiferromagnetic_2009,kofu_neutron_2009,kaneko_columnar_2008,zhao_spin_2008,aczel_muon-spin-relaxation_2008, kitagawa_commensurate_2008, kitagawa_antiferromagnetism_2009, pratt_coexistence_2009, laplace_atomic_2009, bonville_incommensurate_2010, lumsden_magnetism_2010, huang_neutron-diffraction_2008, dai_magnetism_2012}). This problem can be overcome using fast spectroscopies (on the order of femtoseconds) such as photoemission, x-ray emission (XES), and x-ray absorption (XAS) spectroscopy \cite{mannella_magnetic_2014,bondino_evidence_2008,vilmercati_itinerant_2012,gretarsson_revealing_2011,gretarsson_revealing_2011, pelliciari_magnetic_2016} that can take snapshots of the magnetic moment and reveal higher values of the local fluctuating magnetic moment ($\mu_\text{bare}$) with respect to their slower counterparts.
Moreover, a lot of attention has been paid to spin excitations, with inelastic neutron scattering (INS) being at the forefront in the measurement of spin waves in both parent and doped compounds \cite{zhang_effect_2014,carr_electron_2016, harriger_nematic_2011,wang_doping_2013,luo_electron_2013,luo_electron_2012,ramazanoglu_two-dimensional_2013, stewart_superconductivity_2011,johnson_iron-based_2015,hosono_iron-based_2015,fujita_progress_2011,inosov_spin_????,tranquada_superconductivity_2014,chubukov_iron-based_2015,dai_magnetism_2012,dai_antiferromagnetic_2015,lumsden_magnetism_2010}. Recently, resonant inelastic x-ray scattering (RIXS) has also been able to successfully measure high energy spin excitation in parent, electron-, and hole-doped Fe pnictides \cite{zhou_persistent_2013,pelliciari_intralayer_2016,pelliciari_presence_2016} obtaining similar results to INS, and extending the range of experimental techniques available to map out spin excitations. 

$\mathrm{EuFe_2As_2}$ is an interesting compound of the 122 series because of the magnetism present on Eu$^{2+}$. Eu is a rare earth element having \textit{f} electrons that in the $2+$ oxidation state are in a 4\textit{f}$^7$ electronic configuration and consequently Eu has a total angular moment $J=7/2$ that is obtained from $L=0$ and $S=7/2$. Because of this, $\mathrm{EuFe_2As_2}$ has two magnetic transitions, the first one at $\approx$190~K is analogue to $\mathrm{BaFe_2As_2}$ and concerns the AF ordering of the spins at the Fe sites, meanwhile the second one is observed at $\approx$20~K and involves the ordering of spins on Eu$^{2+}$ (Ref.~\cite{ren_antiferromagnetic_2008,xiao_magnetic_2009,jeevan_interplay_2011,nandi_magnetic_2014,herrero-martin_magnetic_2009}). The spins of Eu order in an A-type AF ordering (the ordering is AF between different layers but ferromagnetic inside a single Eu plane) \cite{ren_antiferromagnetic_2008,xiao_magnetic_2009,jeevan_interplay_2011,nandi_magnetic_2014,herrero-martin_magnetic_2009}. This additional magnetic phase and the constant presence of high magnetic moment on the spacing layer do not preclude the emergence of SC when doping is performed \cite{ren_superconductivity_2009,jeevan_electrical_2008,jeevan_high-temperature_2008,jeevan_interplay_2011,miclea_evidence_2009,zapf_$mathrmeufe_2mathrmas_1-xmathbfp_x_2$:_2013,nandi_coexistence_2014, maiwald_signatures_2012, tokiwa_unique_2012}, representing  one of few cases, where ferromagnetism and SC are close in the phase diagram of Fe pnictides \cite{nandi_coexistence_2014,jeevan_interplay_2011}, posing the natural question whether the high magnetic moment of Eu$^{2+}$ affects the magnetism of the FeAs layer. Generally, the bandwidth of the spin excitations has been observed to be affected by the geometry of the FeAs layer \cite{pelliciari_presence_2016,zhang_effect_2014,gretarsson_spin-state_2013}. In $\mathrm{EuFe_2As_2}$ the magnetism on Eu ions is an additional tuning parameter for the magnetism that has to be investigated.

In this article, we report on the detailed characterization of the Fe magnetism in $\mathrm{EuFe_2As_2}$. We have combined hard and soft x-ray spectroscopies to characterize the local magnetic moment and the dispersion of spin excitations. Using XES, we find $\mu_\text{bare}$ to be 1.3$\pm$0.15 at 15~K and to slightly increase to $\mu_\text{bare}=$1.45$\pm$0.15 at 300~K on the Fe atoms. Employing Fe-L$_{2, 3}$ RIXS, we map out the spin excitations as a function of momentum transfer along two high symmetry directions ($\mathrm{(0, 0)\rightarrow(1, 0)}$ and $\mathrm{(0, 0)\rightarrow(1, 1)}$) and extract the dispersion relation. At high momentum transfer the bandwidth of the magnetic excitations is about 180 meV at (0.44, 0) and (0.31, 0.31). We compare and contrast our results on $\mathrm{EuFe_2As_2}$ with $\mathrm{BaFe_2As_2}$ demonstrating that the magnetism in the two related compounds is similar in terms of both $\mu_\text{bare}$ and spin wave dispersion.

\begin{figure}
\includegraphics[scale=0.4]{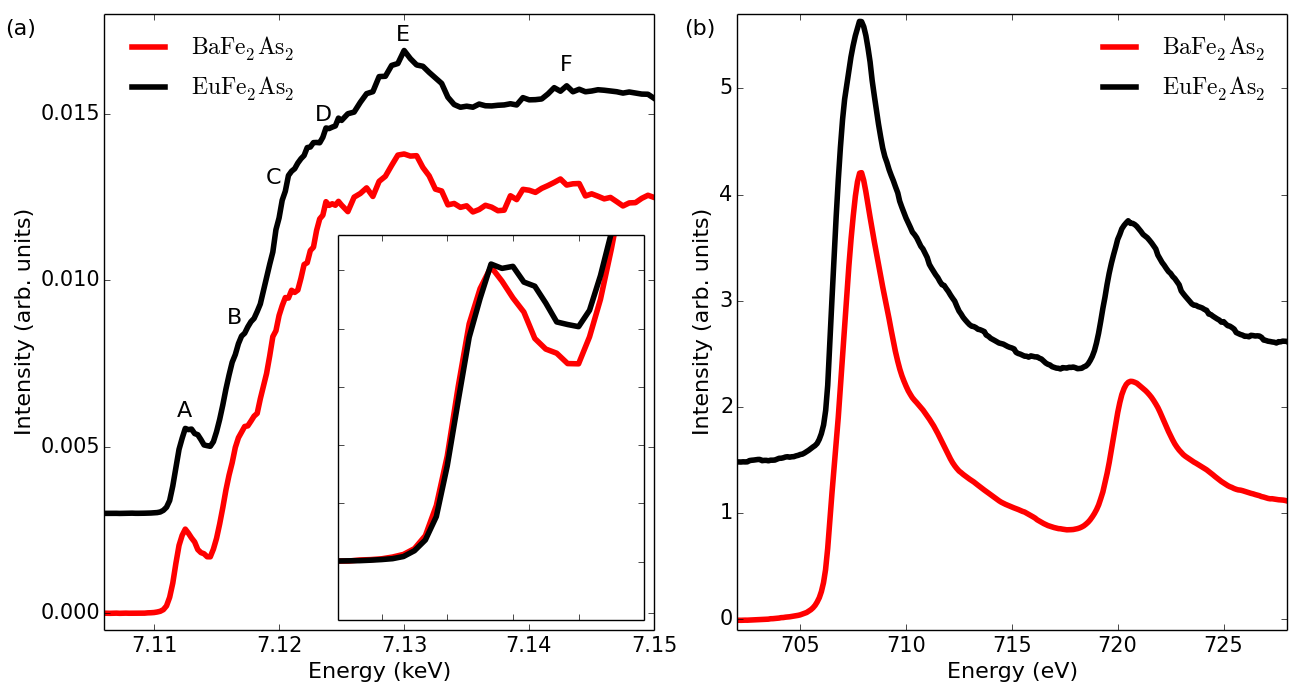}
\caption{\label{fig:fig1} (a) Fe-K XAS-PFY for $\mathrm{EuFe_2As_2}$ and $\mathrm{BaFe_2As_2}$. Inset: Zoom on the pre-edge region. (b) Fe-L$_{2, 3}$ XAS-TFY for $\mathrm{EuFe_2As_2}$ and $\mathrm{BaFe_2As_2}$.}
\end{figure}

\section{Experimental and methods}
\subsection{Sample preparation}
Single crystals were obtained using a Bridgman method following the procedure described in \cite{jeevan_electrical_2008}. Eu  (99.99\%), Fe (99.99\%), and As (99.99999\%) were mixed with a molar ratio of 1:2:2 and put in $\mathrm{Al_2O_3}$ crucible, sealed then in a $\mathrm{Ta}$ crucible under argon atmosphere. The crucible was heated at $30\,^{\circ}\mathrm{C}$ / h to $600\,^{\circ}\mathrm{C}$ and kept there for 12 h and then brought and kept constant at $900\,^{\circ}\mathrm{C}$ for 1 h. After this, the sample is left at $1300\,^{\circ}\mathrm{C}$ for 3 h, and subsequently slowly cooled down. We obtained plate-like crystals whose quality was checked using Laue and scanning electron microscopy equipped with energy dispersive x-ray analysis. 

\subsection{Experimental conditions}
\subsubsection{Fe-K edge XAS and XES}
Fe-K edge XAS and XES experiments were carried out at BL11XU of SPring-8, Hyogo, Japan. The incoming photon beam was monochromatized by a Si(111) double-crystal and a Si(400) secondary channel-cut crystal. We calibrated the energy by measuring x-ray absorption of a polycrystalline Fe foil. We employed three spherical diced Ge(620) analyzers and a detector in Rowland geometry at ca 2 m distance from the analyzers. The total combined resolution was about 400 meV, estimated from the FWHM of the elastic line. XAS at the Fe-K edge was measured in partial fluorescence yield (PFY) mode setting the analyzer to 7.059 keV and scanning the incident energy from 7.10 keV up to 7.15 keV. The intensity was normalized by the incident flux monitored by an ionization chamber. XAS-PFY spectra were all collected at 15~K.
The energy of the incident x-rays of the XES experiments was set to 7.140 keV with $\pi$ polarization and the outgoing photon energy was scanned between 7.02 keV and 7.08 keV. The intensity was normalized as for XAS by the incident flux monitored by an ionization chamber. XES spectra were recorded at 15 and 300~K employing a closed cycle He cryostat.

\subsubsection{Fe-L$_{2,3}$ edge XAS and RIXS}
We post-cleaved the samples \textit{in situ} at a pressure close to 2.0 $\times$ 10$^{-10}$ mbar and mounted them for Fe-L$_{2,3}$ XAS and RIXS experiments with the \textit{ab} plane perpendicular to the scattering plane and the \textit{c} axis lying in it (see the sketch in Fig.~\ref{fig:fig3}(a)). We collected RIXS spectra along the $\mathrm{(0, 0)\rightarrow(1, 0)}$ and $\mathrm{(0, 0)\rightarrow(1, 1)}$ crystallographic directions according to the orthorhombic unfolded notation \cite{park_symmetry_2010}. We express the momentum transfer (q$_{//}$) as relative lattice units (R.L.U.) (q$_{//}\times a/2\pi$) and use the convention of 1 Fe per unit cell. All the measurements were carried out at 19~K by cooling the manipulator with liquid helium. Fe-L$_{2,3}$ XAS and RIXS experiments were performed at the ADRESS beamline of the Swiss Light Source, Paul Scherrer Institute, Villigen PSI, Switzerland \cite{strocov_high-resolution_2010,ghiringhelli_saxes_2006}. XAS spectra at the Fe-L$_{2,3}$ edge were measured in total fluorescence yield (TFY) employing a diode detector. We measured the Fe-L$_{2,3}$~XAS spectra at 15 degrees incidence angle relative to the sample surface. The RIXS spectrometer was set to a scattering angle of 130 degrees and the incidence angle on the sample's surface was varied to change the in-plane q$_{//}$ from (0, 0) to (0.44, 0) and from (0, 0) to (0.31, 0.31) as shown in the sketch of Fig.~\ref{fig:fig3}(a). All RIXS data shown in the current paper are collected at grazing incidence as illustrated in the scheme of Fig.~\ref{fig:fig3}(a). The total energy resolution of the RIXS experiments has been measured employing the elastic scattering of carbon-filled acrylic tape and is around 90 meV.

\section{Results and discussion}
XAS is a very well established technique to characterize the electronic ground states in materials with element selectivity \cite{groot_core_2008,groot_multiplet_2005,de_groot_high-resolution_2001,bittar_co-substitution_2011}.
Figure~\ref{fig:fig1}(a) shows a summary of the Fe-K XAS-PFY spectra acquired on $\mathrm{EuFe_2As_2}$ (black solid line) and $\mathrm{BaFe_2As_2}$ (red solid line) that was also measured for comparison. We can see the main features due to dipole allowed transitions between Fe-1$s$ and Fe-4$p$ states (labeled as peaks B, C, D, E, and F) as well as a weak pre-edge at energy 7.1127 keV (displayed also in the inset of Fig.~\ref{fig:fig1}(a)) in agreement with a previous work for $\mathrm{BaFe_2As_2}$ \cite{bittar_co-substitution_2011}. This pre-edge peak is ascribed to hybridization between Fe-3$d$ and As-4$p$ orbitals, commonly seen in materials lacking inversion symmetry \cite{lafuerza_evidences_2016}. We observe little modification of Fe-K XAS-PFY between $\mathrm{BaFe_2As_2}$ and $\mathrm{EuFe_2As_2}$ as shown in the inset illustrating the pre-edge region, where we can see just a small difference in the width of the peak A. This small modification indicates that the hybridization and covalency between Fe and As is very similar in these compounds and marginally affected by the spacing layer. A slight change in the relative intensity of the peaks D and F is observed in Fig.~\ref{fig:fig1}(a). These small effects may be ascribed to the local structural differences between the two systems originating from the different spacing layer, slightly changing the structure but not the hybridization.
In Fig.~\ref{fig:fig1}(b), we plot the Fe-L$_{2, 3}$ XAS spectrum of $\mathrm{EuFe_2As_2}$ together with the one of $\mathrm{BaFe_2As_2}$. The spectra display two broad peaks at $\approx$708 and 720 eV typical of Fe pnictides, ascribed to the L$_3$ and L$_2$ absorption edges \cite{zhou_persistent_2013,pelliciari_intralayer_2016,pelliciari_presence_2016,kurmaev_identifying_2009,hancock_evidence_2010}. In both compounds, the spectra are rather similar with minor differences in the shoulder at 711 eV. Usually this shoulder involves the charge transfer satellites and the small difference observed in the Fe-L XAS may be the effect of the different spacing layer affecting the distance between Fe and As, and consequently changing the charge transfer peak, as also suggested by Fe-K edge XAS-PFY. Clearly, Fe-L edge XAS confirms that in $\mathrm{EuFe_2As_2}$ Fe has the common $2+$ oxidation state. Furthermore, the spectra do not show any multiplet structure common to oxides, confirming the high sample quality and the success of the \textit{in situ} cleaving \cite{zhou_persistent_2013,pelliciari_presence_2016, pelliciari_intralayer_2016,kurmaev_identifying_2009,hancock_evidence_2010}. The FWHM of this peak is about 3 eV and is unchanged moving from $\mathrm{EuFe_2As_2}$ to $\mathrm{BaFe_2As_2}$, indicating that the electronic structure is similar for the two compounds.

XES is a powerful probe for the measurement of $\mu_\text{bare}$ \cite{bergmann_x-ray_2009, vanko_probing_2006, gretarsson_spin-state_2013, gretarsson_revealing_2011, ortenzi_structural_2015, simonelli_coexistence_2012, simonelli_temperature_2014,pelliciari_magnetic_2016}. One excites the Fe 1$s$ core-electron into the continuum by means of a photon (in our case $h\nu$=7.140 keV) creating a highly unstable core-hole. This core-hole can be filled by an Fe 3$p$ electron with the consequent emission of a photon to satisfy the energy balance. The system is then left in Fe 3$p^5$ final state which has a wavefunction affected by the Fe 3$d$ orbitals and sensitive to the spin carried by valence electrons. The XES line created in this process is the Fe-K$_{\beta}$ emission line that is composed by a main peak, due to the sum of K$_{\beta_{1}}$ and K$_{\beta_{3}}$, and a satellite peak named K$_{\beta'}$. The latter has been shown to be directly sensitive to the spin of the valence band, and using a proper calibration it is possible to extract the value of $\mu_\text{bare}$ \cite{bergmann_x-ray_2009, vanko_probing_2006, gretarsson_spin-state_2013, gretarsson_revealing_2011, ortenzi_structural_2015, simonelli_coexistence_2012, simonelli_temperature_2014,pelliciari_magnetic_2016,glatzel_high_2005,lafuerza_evidences_2016}. This technique is able to probe $\mu_\text{bare}$ in the femtosecond range overcoming the drawback of the quenching of the magnetic moment due to quantum fluctuations concerning other probes \cite{mannella_magnetic_2014}.

\begin{figure}
\includegraphics[scale=0.4]{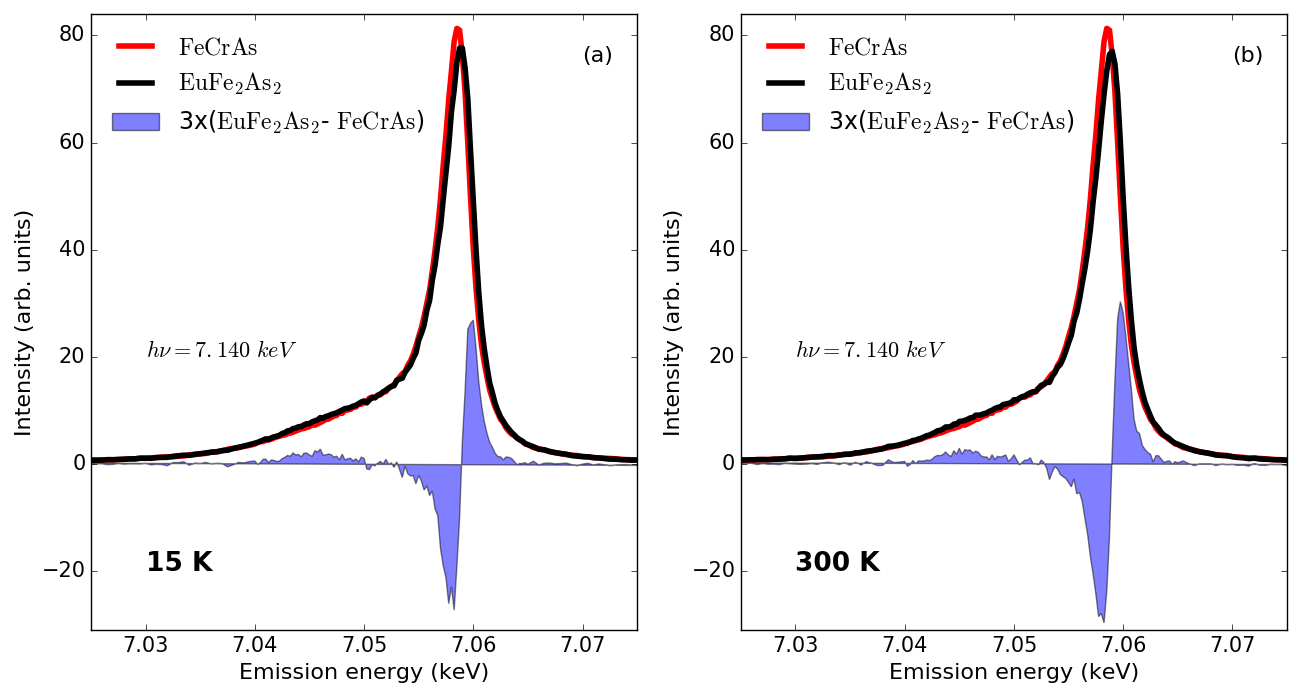}
\caption{\label{fig:fig2} (a) Fe-K$_\beta$ XES for $\mathrm{EuFe_2As_2}$ and FeCrAs (employed for calibration) at 15~K. The shadowed part indicates the difference spectrum for $\mathrm{EuFe_2As_2}$ from which the IAD is extracted. (b) Fe-K$_\beta$ XES for $\mathrm{EuFe_2As_2}$ and FeCrAs (employed for calibration) at 300~K. The shadowed part indicates the difference spectrum for $\mathrm{EuFe_2As_2}$ from which the IAD is extracted.}
\end{figure}

We show XES spectra in Fig.~\ref{fig:fig2}(a) and (b) for $\mathrm{EuFe_2As_2}$ as solid black lines at 15 and 300~K, respectively. As red solid line we plot also the reference spectrum collected on FeCrAs, which has been chosen, as in previous works \cite{gretarsson_revealing_2011,gretarsson_spin-state_2013,pelliciari_magnetic_2016,simonelli_temperature_2014}, because of the absence of magnetic moment on Fe atoms and represents a suitable reference material to apply the integrated absolute difference (IAD) method \cite{gretarsson_revealing_2011,gretarsson_spin-state_2013,pelliciari_magnetic_2016,simonelli_temperature_2014}. Additionally in Fig.~\ref{fig:fig2}(a) and (b) we illustrate as blue shadowed region the difference spectra obtained after using the same center of mass and subtraction of the reference spectrum as described in previous works \cite{gretarsson_revealing_2011,gretarsson_spin-state_2013,pelliciari_magnetic_2016,simonelli_temperature_2014,glatzel_high_2005,bergmann_x-ray_2009,lafuerza_evidences_2016}. This procedure allows to determine the value of $\mu_\text{bare}=1.3\pm0.15$ for $\mathrm{EuFe_2As_2}$ at 15~K and 1.45$\pm0.15$ at 300~K. The values measured for $\mathrm{EuFe_2As_2}$ are similar to the values obtained for $\mathrm{BaFe_2As_2}$ ($\mu_\text{bare}=1.3\pm0.15$ at 15~K and $\mu_\text{bare}=1.5\pm0.15$ at 300~K) \cite{gretarsson_revealing_2011,pelliciari_magnetic_2016}. Similarly, the small increase of $\mu_\text{bare}$ with temperature is consistent with what is observed in other Fe pnictides \cite{gretarsson_revealing_2011,gretarsson_spin-state_2013,pelliciari_magnetic_2016} and demonstrates that the use of Fermi surface nesting arguments as the sole source of magnetic ordering in $\mathrm{EuFe_2As_2}$ is inappropriate \cite{pelliciari_magnetic_2016,richard_fe-based_2011, mazin_superconductivity_2010, stewart_superconductivity_2011, johnston_puzzle_2010, graser_near-degeneracy_2009}. The increase of $\mu_\text{bare}$ with temperature is incompatible with the nesting scenario and it has been shown that instead a spin freezing transition driven by the competition of Hund's coupling, electronic screening, and Fermi surface nesting arguments is more likely and in agreement with systematic spectroscopic measurements on parent and doped Fe pnictides \cite{pelliciari_magnetic_2016,werner_spin_2008,werner_satellites_2012,gretarsson_spin-state_2013,tytarenko_direct_2015,chaloupka_spin-state_2013}.

\begin{figure}
\centering
\includegraphics[scale=0.4]{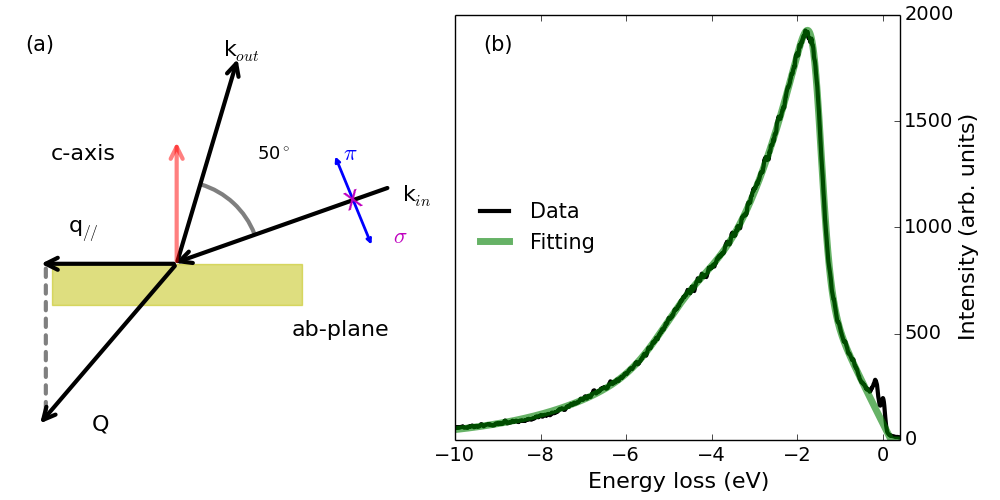}
\caption{\label{fig:fig3} (a) Experimental setup for RIXS experiment. (b) Black solid line: RIXS spectrum for $\mathrm{EuFe_2As_2}$ at 19~K at (0.44, 0). Green solid line: Fitting of the emission line.}
\end{figure}

In recent years, RIXS has received a lot of attention due to the huge instrumental improvements in terms of both energy resolution and flux \cite{strocov_high-resolution_2010,ghiringhelli_saxes_2006} permitting the detection of magnetic excitations in high temperature superconductors such as cuprates and Fe pnictides \cite{braicovich_dispersion_2009,braicovich_momentum_2010,braicovich_magnetic_2010,dean_high-energy_2013,dean_itinerant_2014,dean_spin_2012,dean_persistence_2013,lee_asymmetry_2014,zhou_persistent_2013,pelliciari_intralayer_2016,wakimoto_high-energy_2015,ishii_high-energy_2014,peng_magnetic_2015,minola_collective_2015,ellis_correlation_2015,dean_insights_2015,monney_resonant_2016,pelliciari_presence_2016,dantz_quenched_2016}.
We measured RIXS on $\mathrm{EuFe_2As_2}$ on the same single crystal used for XES and XAS experiments. The experimental configuration is shown in Fig.~\ref{fig:fig3}(a). Figure~\ref{fig:fig3}(b) displays as a black solid line a RIXS spectrum collected at (0.44, 0) and 19~K with the incident energy set to the maximum of the Fe-L$_{2, 3}$ XAS peak ($\approx$708 eV). The spectrum has a main peak at around -2 eV in energy loss and an additional shoulder at -4.3 eV. It is rather similar to the resonant emission observed in intermetallic materials of Fe and As \cite{zhou_persistent_2013,pelliciari_intralayer_2016,pelliciari_presence_2016,hancock_evidence_2010,kurmaev_identifying_2009,yang_evidence_2009}. Despite the cleaving and high data quality, we do not see \textit{dd}-excitations observed in other Fe pnictides with soft and hard x-ray RIXS \cite{gretarsson_resonant_2015,nomura_resonant_2016}. In Fig.~\ref{fig:fig3}(b) we additionally show the fitting of the full emission line carried out as in Refs.~\cite{pelliciari_intralayer_2016,pelliciari_presence_2016,zhou_persistent_2013,hancock_evidence_2010} employing the following formulas:

\begin{equation}
I_{fit} = (\beta x^{2} + \alpha x + c )  \cdot(1 - g_{\gamma}) + I_{0}\exp(ax) \cdot g_{\gamma} + G \nonumber 
\end{equation} 
\textrm{with}\\

\begin{equation}
g_{\gamma} =  \left(\exp \left(\frac{x + \omega^{\ast}} {\Gamma}\right) + 1\right)^{-1} \nonumber 
\end{equation} 
\textrm{and}\\
\begin{equation}
G = \frac{A}{\sigma \sqrt{2\pi}} \exp\left(\frac{(x + x_{0})^2} {2\sigma^2}\right) \nonumber 
\end{equation} 

where the low energy loss region (x) is fitted with a second order polynomial (with $\alpha,$ $\beta$ and $c$ as parameters) that is transited into an exponential decay (defined by intensity $I_{0}$ and slope $a$) at higher energy loss. This crossover between the two functions is obtained employing the function g$_\gamma$ that includes the energy $\omega^{\ast}$ and the width $\gamma$ of such a crossover \cite{hancock_evidence_2010,pelliciari_intralayer_2016}. The shoulder visible at energy loss around -4.3 eV is fitted with a gaussian term $G$.

\begin{figure}
\centering
\includegraphics[width=\textwidth]{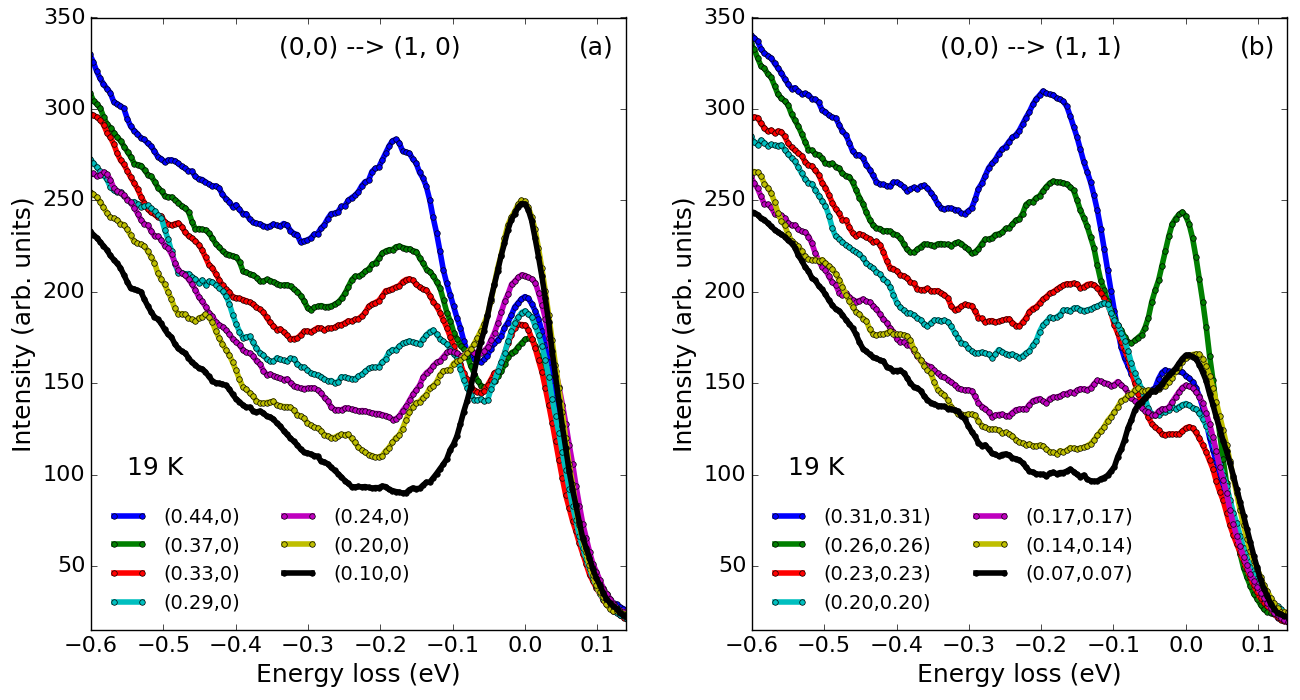}
\caption{\label{fig:fig4} (a) $\mathrm{EuFe_2As_2}$~RIXS spectra at 19~K for different q$_{//}$ along the $\mathrm{(0, 0)\rightarrow(1, 0)}$ direction. The q$_{//}$ values are (0.44, 0), (0.37, 0), (0.33, 0), (0.29, 0), (0.24, 0), (0.20, 0), and (0.10, 0). (b) RIXS spectra at 19~K and different q$_{//}$ along the $\mathrm{(0, 0)\rightarrow(1, 1)}$ direction. The q$_{//}$ values are (0.31, 0.31), (0.26, 0.26), (0.23, 0.23), (0.20, 0.20), (0.17, 0.17), (0.14, 0.14), and (0.07, 0.07).}
\end{figure}

\begin{figure}
\centering
\includegraphics[scale=0.35]{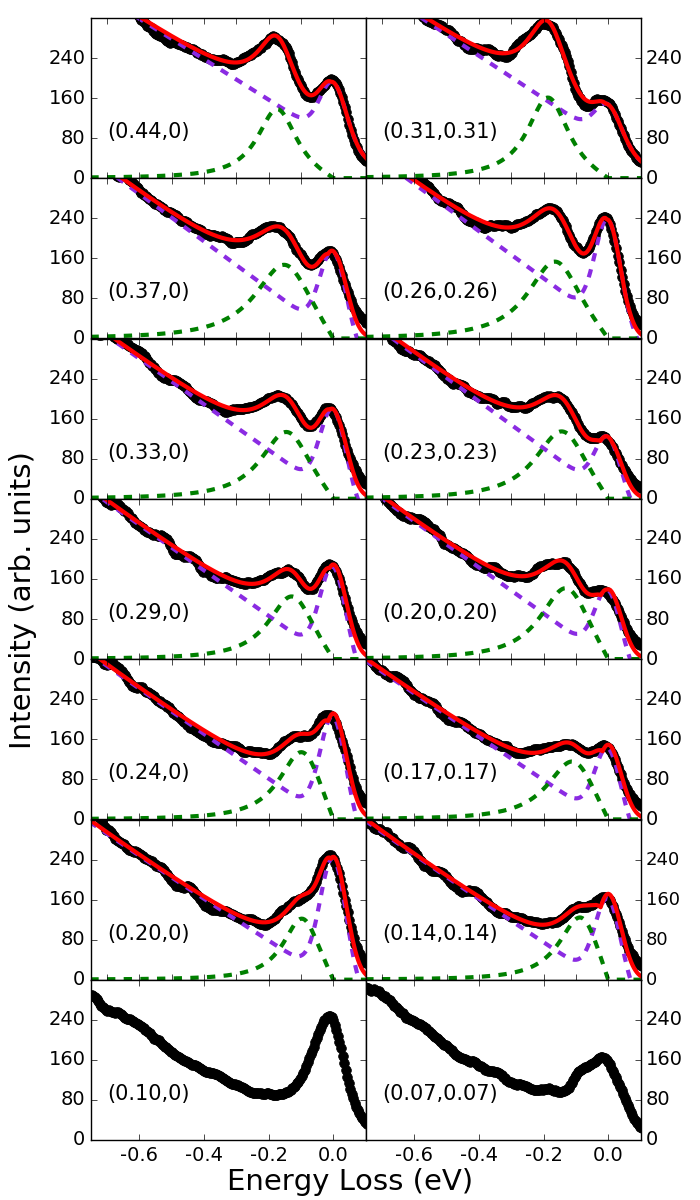}
\caption{\label{fig:fig4fit} RIXS spectra of $\mathrm{EuFe_2As_2}$ at 19~K and different momentum transfer along (0, 0)$\rightarrow$(1, 0) (left panel) and (0, 0)$\rightarrow$(1, 1) (right panel). Black dots are experimental data, solid purple is the sum of resonant emission and elastic, green is the spin excitation peak and red is the sum of all the components of the fitting. At (0.10, 0) and (0.07, 0.07) we do not attempt any fitting since a peak for the magnetic excitations is not clearly visible.}
\end{figure}

\begin{figure}
\centering
\includegraphics[scale=0.4]{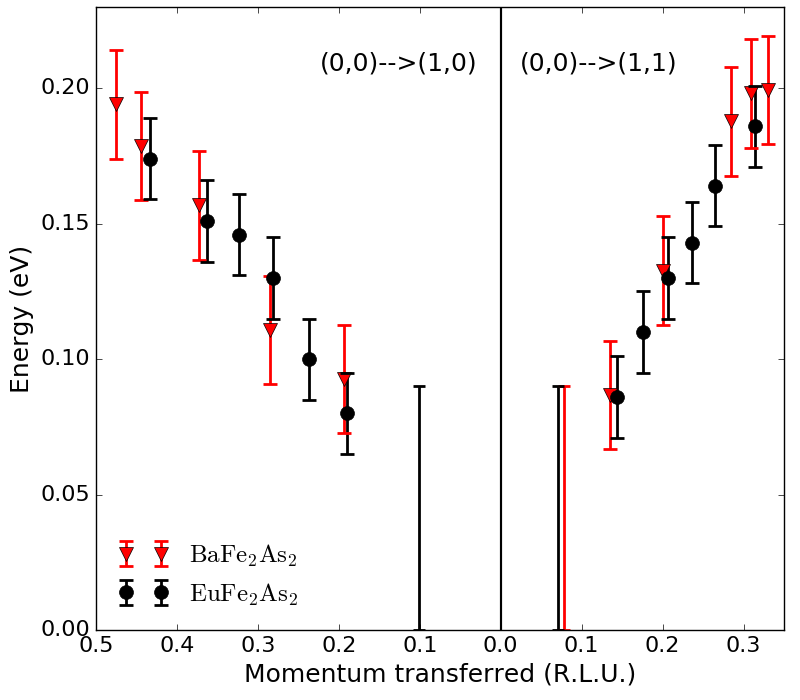}
\caption{\label{fig:fig5} Dispersion of magnetic excitations of $\mathrm{EuFe_2As_2}$ at 19~K along (0, 0)$\rightarrow$(1, 0) and (0, 0)$\rightarrow$(1, 1) directions as black dots with error bars. The dispersion of $\mathrm{BaFe_2As_2}$ is shown as red triangles with error bars for comparison purpose as extracted from Ref.~\cite{zhou_persistent_2013}. At (0.1, 0) and (0.07, 0.07) the fitting was not possible because of overlap of the magnetic peak with the elastic line. Only an estimation of the energy range is provided for these values, as depicted by error bars. The points have been slightly shifted along the horizontal axis for better visualization.}
\end{figure}

At high q$_{//}$ in the low energy loss region (between 0 and -0.6 eV in energy loss) a clear peak well separated from the elastic line appears as shown in Fig.~\ref{fig:fig4}(a) for the $\mathrm{(0, 0)\rightarrow(1, 0)}$ directions and Fig.~\ref{fig:fig4}(b) for the $\mathrm{(0, 0)\rightarrow(1, 1)}$ direction. We change the incident angle in order to tune q$_{//}$ to the following values depicted as different colored lines with dots: (0.44, 0), (0.37, 0), (0.33, 0), (0.29, 0), (0.24, 0), (0.20, 0), and (0.10, 0) in Fig.~\ref{fig:fig4}(a) and (0.31, 0.31), (0.26, 0.26), (0.23, 0.23), (0.20, 0.20), (0.17, 0.17), (0.14, 0.14), and (0.07, 0.07) in Fig.~\ref{fig:fig4}(b). 
The peak appearing at 180 meV at high q$_{//}$ disperses to lower energy when q$_{//}$ is decreased as displayed in Fig.~\ref{fig:fig4}(a), (b), and Fig.~\ref{fig:fig4fit}. The bandwidth and collective behavior resemble what was previously observed for spin excitations by RIXS experiments on parent $\mathrm{BaFe_2As_2}$, $\mathrm{SmFeAsO}$, and $\mathrm{NaFeAs}$ \cite{zhou_persistent_2013,pelliciari_intralayer_2016,pelliciari_presence_2016}. The low energy fitting analysis (Fig.~\ref{fig:fig4fit}) has been carried out similarly to previous works \cite{zhou_persistent_2013,pelliciari_intralayer_2016,pelliciari_presence_2016} using a gaussian curve for the elastic peak and an anti-symmetrized lorentzian curve for the spin excitation peak as summarized in Fig~\ref{fig:fig4fit}.
From this fitting analysis, we extracted the magnetic dispersion curve outlined in Fig.~\ref{fig:fig5} as black dots with error bars. The spin excitations along the $\mathrm{(0, 0)\rightarrow(1, 0)}$ direction disperses from 170 meV energy loss at (0.44, 0) (limit set by experimental geometry) to below the detection limit (estimated at around 80 meV) close to the $\Gamma$ point. In the $\mathrm{(0, 0)\rightarrow(1, 1)}$ direction the magnetic dispersion is much steeper resulting in an energy of 180 meV at (0.31, 0.31). As red dots with error bars in Fig.~\ref{fig:fig5}, we overplot for comparison the spin excitation dispersion of $\mathrm{BaFe_2As_2}$ extracted from Ref.~\cite{zhou_persistent_2013}. The spin excitations spectrum for the two compounds is very similar at this temperature, indicating little effects due to the presence of the magnetism on Eu. It was proposed in Ref.~\cite{jeevan_electrical_2008} that the two magnetic orderings are decoupled. From our data, the lack of changes revealed in the spin excitations spectrum of $\mathrm{EuFe_2As_2}$ and $\mathrm{BaFe_2As_2}$ as well as the similar values of $\mu_\text{bare}$ are in complete agreement with this hypothesis. Alternatively, if there is any effect and coupling arising from Eu our experimental probes are not sensitive enough to detect it. However, further studies performed at lower temperature and in presence of magnetic field might unravel additional effects.

\begin{figure}
\centering
\includegraphics[scale=0.55]{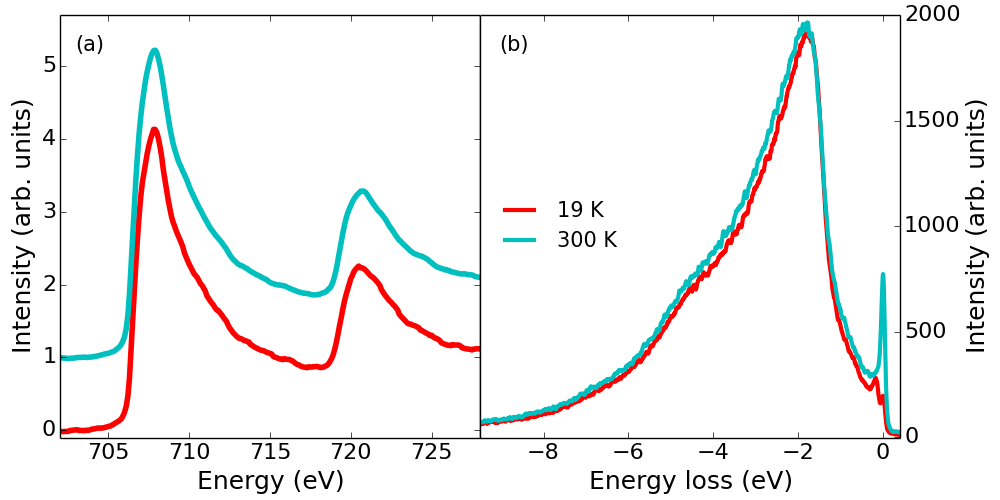}
\caption{\label{fig:fig6} (a) $\mathrm{EuFe_2As_2}$ Fe-L$_{2, 3}$ XAS at 19~K (red) and 300~K (cyan). (b) RIXS spectra at (0.44, 0), 19~K (red) and 300~K (cyan) for $\mathrm{EuFe_2As_2}$.}
\end{figure}

\begin{figure}
\centering
\includegraphics[scale=0.4]{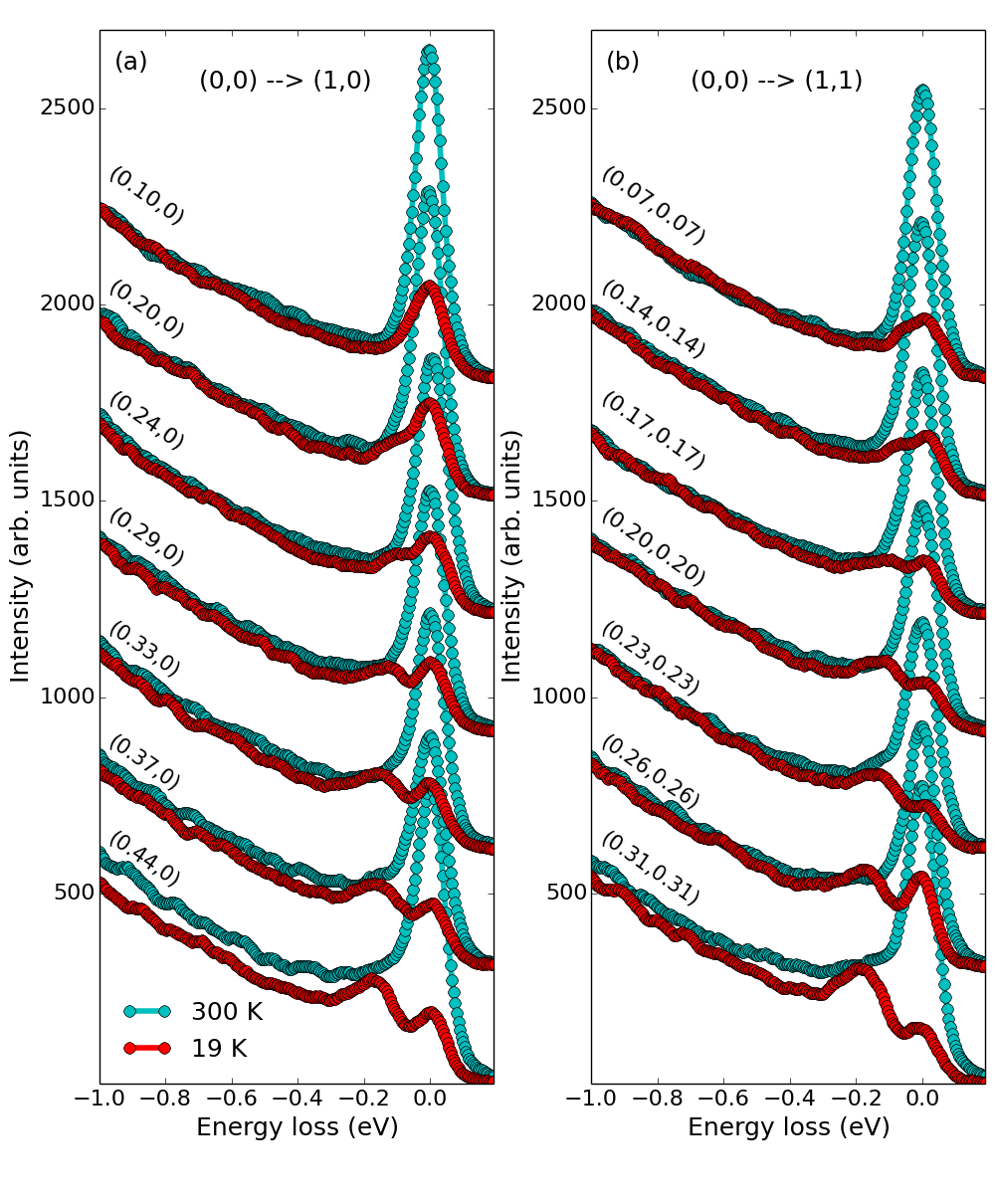}
\caption{\label{fig:fig7} (a) Comparison between RIXS spectra for $\mathrm{EuFe_2As_2}$ at 19~K (red) and 300~K (cyan) along the $\mathrm{(0, 0)\rightarrow(1, 0)}$  direction. (b) Comparison between RIXS spectra for $\mathrm{EuFe_2As_2}$ at 19~K (red) and 300~K (cyan) along the $\mathrm{(0, 0)\rightarrow(1, 1)}$  direction.}
\end{figure}

In addition, we performed the same XAS and RIXS experiments at 300~K. Figure~\ref{fig:fig6}(a) illustrates the Fe-L$_{2, 3}$ XAS at 19 and 300~K. The Fe-L$_{2, 3}$ XAS spectra are invariant with temperature as visible in Fig.~\ref{fig:fig6}(a). In analogy to the measurements carried out at 19~K, we tuned the incoming energy to the maximum of the Fe-L$_{3}$ XAS and acquired RIXS spectra at the same incident angles. In Fig.~\ref{fig:fig6}(b) we show a comparison between RIXS spectra recorded at (0.44, 0) and 19~K (red solid line) as well as 300~K (cyan solid line). The spectra have been normalized to the acquisition time, while ensuring that the incident photon flux on the sample was constant (monitored through the measurement of the drain current on the last optical element before the sample). The main emission line is slightly broadened by the increased temperature especially in the region between -2.1 eV and -4.3 eV, due the effect of the phase transition and the consequent change of the band structure. Additionally, as clearly seen in Fig.~\ref{fig:fig6}(b) and~\ref{fig:fig7}(a-b), the elastic line is hugely increased at high temperature. The lack of any Fe$^{3+}$ peak in the Fe-L$_{2, 3}$ XAS tells us that the sample has not been damaged by raising the temperature, suggesting that the increase of the elastic line is rather an intrinsic property of the sample. This can, in principle, be ascribed to several phenomena such as an increase in thermal diffuse scattering involving both phonon population and an enhanced scattering rate of electrons at 300~K. We are more inclined to ascribe the growth of the elastic line to the latter cause, since it is a common phenomenon observed in metals even if a possible enhancement arising from thermal diffuse scattering can not be completely ruled out (for details see Supplementary material). In Fig.~\ref{fig:fig7}(a) and (b), the full momentum dependence at 300~K along both $\mathrm{(0, 0)\rightarrow(1, 0)}$ and $\mathrm{(0, 0)\rightarrow(1, 1)}$ is plotted as cyan lines with dots together with the corresponding spectra at low temperature depicted as red lines with dots. The spectral weight in the low energy region (between -50 meV and -300 meV in energy loss) is similar between the two temperatures, however at high temperature a clear peak is not observed. Unfortunately, the data presented here are not conclusive to establish whether magnetic excitations persist at all or not at 300~K in $\mathrm{EuFe_2As_2}$. In Ref.~\cite{zhou_persistent_2013} the effect of the temperature on the RIXS spectra has been shown for $\mathrm{BaFe_2As_2}$ and $\mathrm{Ba_{0.6}K_{0.4}Fe_2As_2}$ for q=0.5 \AA$^{-1}$. Similarly to our results in their case an enhancement in the elastic/quasielastic line was observed when the temperature is raised, demonstrating that this can be a common feature between different families of Fe pnictides. However, the lack of additional momentum points in their study limits the experimental evidence in $\mathrm{BaFe_2As_2}$ and $\mathrm{Ba_{0.6}K_{0.4}Fe_2As_2}$.
We believe that high temperature RIXS studies deserve further attention to understand the evolution of spin excitations and the intrinsic enhancement of the elastic line. In this context, higher resolution instrumentation being commissioned in the future can help to shed light on this behavior by decreasing the width at the base of the elastic line, possibly resolving the spin excitation peak together with other additional excitations such as phonons.

\section{Conclusions}
In summary, we have measured XAS at the Fe-K and L$_{2, 3}$ edge of $\mathrm{EuFe_2As_2}$ and observed that Fe in this compound has a similar electronic state to $\mathrm{BaFe_2As_2}$. XES measurements of the local magnetic moment evidenced a magnetic moment of 1.3$\pm0.15~\mu_B$ at 15~K that is slightly increased to 1.45$\pm0.15~\mu_B$ at 300~K. RIXS experiments detected collective spin excitations along $\mathrm{(0, 0)\rightarrow(1, 0)}$ and $\mathrm{(0, 0)\rightarrow(1, 1)}$ with a bandwidth around 170-180 meV at (0.44, 0) and (0.31, 0.31). We collected momentum dependent RIXS spectra at 300~K and observed a strong increase of the elastic line which precluded any further analysis regarding the spin excitations at this temperature. 
From our observations we conclude that the magnetism of $\mathrm{EuFe_2As_2}$ is similar to $\mathrm{BaFe_2As_2}$ in terms of both $\mu_\text{bare}$ and bandwidth of the spin excitations. This demonstrates that, if there is an effect of the magnetism of Eu$^{2+}$ on the magnitude of the local and correlated magnetism of the FeAs layers, it should be small.

\begin{acknowledgments}
We acknowledge Jannis Maiwald for discussions and carefully reading the manuscript.
J.P. and T.S. acknowledge financial support through the Dysenos AG by Kabelwerke Brugg AG Holding, Fachhochschule Nordwestschweiz, and the Paul Scherrer Institut. J. P. acknowledge financial support by the Swiss National Science Foundation Early Postdoc.Mobility fellowship project number P2FRP2$\_$171824. The synchrotron radiation experiments have been performed at BL11XU of SPring-8 with the approval of the Japan Synchrotron Radiation Research Institute (JASRI) (Proposals No. 2014A3502 and 2014B3502) and at the ADRESS beamline of the Swiss Light Source at the Paul Scherrer Institut. We thank D. Casa for fabrication of the Ge(620) analyzers installed at BL11XU of SPring-8. Part of this research has been funded by the Swiss National Science Foundation through the Sinergia network Mott Physics Beyond the Heisenberg (MPBH) model, the D-A-CH program (SNSF Research Grant No. 200021L 141325), and by the NCCR MARVEL, funded by the Swiss National Science Foundation. The research leading to these results has received funding from the European Community’s Seventh Framework Programme (FP7/2007-2013) under Grant Agreement No. 290605 (COFUND: PSIFELLOW). H. S. J. and P. G. acknowledge financial support through the SPP 1458 of the Deutsche Forschungsgemeinschaft.

\end{acknowledgments}

\end{document}